\documentstyle[12pt]{article}
\input epsf

\newlength{\extraspace}
\newlength{\extraspaces}
\newcommand{\be}{\begin{equation}
\addtolength{\abovedisplayskip}{\extraspaces}
\addtolength{\belowdisplayskip}{\extraspaces}
\addtolength{\abovedisplayshortskip}{\extraspace}
\addtolength{\belowdisplayshortskip}{\extraspace}}
\newcommand{\ee}{\end{equation}}

\newcommand{\bq}{\begin{eqnarray}
\addtolength{\abovedisplayskip}{\extraspaces}
\addtolength{\belowdisplayskip}{\extraspaces}
\addtolength{\abovedisplayshortskip}{\extraspace}
\addtolength{\belowdisplayshortskip}{\extraspace}}
\newcommand{\eq}{\end{eqnarray}}

\newcommand{\newsection}[1]{
\vspace{15mm}
\pagebreak[3]
\addtocounter{section}{1}
\setcounter{equation}{0}
\setcounter{subsection}{0}
\setcounter{footnote}{0}
\begin{flushleft}
{\large\bf \thesection. #1}
\end{flushleft}
\nopagebreak
\medskip
\nopagebreak}

\newcommand{\newsubsection}[1]{
\vspace{1cm}
\pagebreak[3]

\addtocounter{subsection}{1}
\noindent{ \bf \thesubsection. #1}
\nopagebreak
\vspace{2mm}
\nopagebreak}

\begin{document}
\hbox{}
\nopagebreak
\vspace{-3cm}
\addtolength{\baselineskip}{.8mm}
\baselineskip=24pt
\begin{flushright}
{\sc OUTP}- 97  65 P\\
hep-th@xxx/9711189 \\
\today
\end{flushright}

\begin{center}
{\Large  Equivalence of the $\beta$-function of the Variational
Approach to that of QCD.}\\
\vspace{0.1in}

{\large William E. Brown}\\
{\it  Theoretical
 Physics,
1 Keble Road, Oxford, OX1 3NP, UK}\\
\vspace{0.1in}
\vspace{0.1in}

{\sc  Abstract}

\end{center}

\noindent
The variational ansatz for the ground state wavefunctional of QCD is
found to capture the anti-screening behaviour that contributes the
dominant `$-4$' to the $\beta$-function and leads to asymptotic freedom.
By considering an SU(N) purely gauge theory in the Hamiltonian
formalism and choosing the Coulomb gauge, the origins of all screening
and anti-screening contributions in gluon processes are found in terms
of the physical degrees of freedom.  The overwhelming anti-screening
contribution of `$-4$' is seen to originate in the renormalisation of a
Coulomb interaction by a transverse gluon.  The lesser screening
contribution of `$+\frac{1}{3}$' is seen to originate in processes
involving transverse gluon interactions.  It is thus apparent how the
variational ansatz must be developed to capture the full running of
the QCD coupling constant.

\noindent

\vfill

\newpage
\newsection{Introduction.}

Asymptotic freedom was discovered in some seminal calculations
published in the 1970's.  This unexpected result was proved for all
non-abelian gauge theories and Quantum Chromodynamics (QCD) was born soon
after.  Khriplovich, \cite{1}, calculated the full Green's functions
of an SU(2) purely gauge theory in the radiation gauge using the
spectral representation.  Within the full Coulomb Green's function of
\cite{1} was written the now well known components of the
$\beta-$function of the SU(2) Yang-Mills theory.  The calculation of
what is now the QCD
$\beta-$function was carried out some years later, \cite{2}, and also
by 't Hooft (unpublished).

The $\beta-$function for a purely gauge field Yang-Mills theory (no
fermions) is,
\begin{equation}
\beta(g) = -\frac{g^3 C_2(G)}{(4\pi)^2}\left[4-\frac{1}{3}\right].
\label{i1}
\end{equation}
The contribution `-$4$' is an overwhelming anti-screening effect that
gives asymptotic freedom.  The `$\frac{1}{3}$' is a lesser screening
effect.  $C_2(G)$ is the second Casimir operator of the SU(N) gauge
group in the adjoint representation.

After the work of \cite{1} and \cite{2}, some attempts were made to
develop an intuitive, physical understanding of the mechanisms by which
the screening and anti-screening effects are manifest.  In \cite{3},
the screening and anti-screening
components of the $\beta-$function were related to the spin of the
gauge field.  The anti-screening effect was related to the
influence of the background field upon the electric dipole density and
the screening effect was related to the magnetic dipole density.  The
$\beta-$function for a field of spin S was written,
\begin{equation}
\beta_S(g) =
-g^3\frac{(-1)^{2S}}{(4\pi)^2}\left[(2S)^2-\frac{1}{3}\right]
\label{i2}
\end{equation}
where the group factor has been omitted.  For a gauge field of spin 1,
the anti-screening effect is $-4$, as in (\ref{i1}).  

The $\beta-$function was decomposed
in \cite{4} into the same screening and anti-screening components as
for a spin 1 particle in \cite{3}, i.e. (\ref{i1}), but with different origins.  This was achieved from the
calculation of the pre-exponential factor, or the renormalisation of
the charge, for the BPST instanton within the path integral formalism.
Expanding the action in terms of a deviation from the instanton field,
the zero-frequency modes were shown to give the anti-screening
contribution of `$-4$' and the positive frequency modes the screening
contribution of `$\frac{1}{3}$'.  There is no contradiction in the
apparently different origins of the effects in these calculations
since the $\beta-$function can be calculated from any physical process
and in \cite{4} is related to instantons.  

For a fuller discussion of the calculations \cite{3} and \cite{4}, the
reader is directed to the originals or to \cite{8} for a brief overview and
discussion in context of the calculation of the $\beta-$function of
the coupling constant of the variational approach.  It is interesting
to note here, however, that the two calculations, which involve physically
different phenomena, produce answers for the $\beta$-function that
decompose identically into screening and anti-screening contibutions.
In this paper, the origins of the same screening and anti-screening
components will be found in terms of fundamental gluon processes.
These gluon process are written in the Hamiltonian formalism in terms
of the physical degrees of
freedom only; transverse gluons and Coulomb interactions.

The corollary of asymptotic freedom is that at low energy there is a
strong coupling problem, e.g. phenomena such as confinement and
chiral symmetry breaking.  After a quarter of a century of work there
is still no complete theory in answer to these questions.  Despite the
suggestion of many promising ideas, the arsenal of necessarily
non-perturbative methods with which to attack these problems is
limited.  To have an analytic solution for the ground state
wavefunctional of an asymptotically free non-abelian gauge theory,
with the associated enhanced understanding of the underlying physics,
would be invaluable for the understanding of these strong coupling
phenomena.  This was the motivation that drove the development of a
variational approach to Quantum Field Theory (QFT).  Although in quantum
mechanical problems the variational approach is often easy to use - a
few qualitative features is usually enough to write an ansatz that
will give good results for the ground state expectation values - there
is rather more complexity to overcome to use the approach within
QFT, as discussed earlier by Feynman, \cite{5}.

The variational approach was successfully applied to QFT in the
exploratory calculations of \cite{6} and \cite{7} where QCD and
Quantum Electrodynamics in $2+1$ dimensions (QED$_3$) were considered,
respectively.  Extensive calculations were performed reproducing many
old and giving new results.  The reader is directed to the original
papers for details.  With the confidence given by the many results
obtained from the variational approach to QCD and QED$_3$, it was
necessary to analyze the ansatz for the ground state wavefunctional of
QCD.  In \cite{6} it was found that the solution of the minimization
equation obtained in the variational approach results in the
variational parameter being fixed away from the perturbative value and
that the best variational state is characterized by a dynamically
generated mass scale.  The vacuum condensate was also calculated.  Hence, the ground state wavefunctional captures
some known non-perturbative characteristics of the QCD vacuum and we
would fully expect the $\beta$-function of the coupling constant
written in the variational ansatz to be like that of QCD.  In fact, it was
assumed in \cite{6} that it was indeed the $\beta$-function of QCD.
This assumption was investigated in \cite{8}.

The calculation of the $\beta-$function from the ansatz for the ground
state wavefunctional of the variational approach was performed in
\cite{8} with the following result.
\begin{equation}
\beta(g) = -4\frac{g^3 C_2(G)}{(4\pi)^2}
\label{i3}
\end{equation}
This result is very close to the $\beta-$function of QCD, (\ref{i1}).
The overwhelming anti-screening contribution that leads to asymptotic
freedom is captured by the variational ansatz.  The lesser screening
contribution is, however, missing.  The effective action, (\ref{v9}),
considered in \cite{8} was written in terms of elements of the SU(N)
gauge group.  Gauss' law, which is the only physical constaint imposed
upon the ground state wavefunctional gives rise to longitudinal gluons
with Coulomb interaction and can be used as the generators of the
group.  So, in \cite{8} it was conjectured that the overwhelming
anti-screening contribution of `$-4$' is due to the renormalisation of
a Coulomb interaction.  Further, since all terms greater than
quadratic were omitted from the
Hamiltonian because they gave only small, $O(g)$, contributions to the
minimization equation, it was conjectured that the screening
contribution `$\frac{1}{3}$' must originate in the interaction of
transverse gluons.  The aim of the calculations presented in this
paper is to prove this conjecture to be correct and so to obtain an
intuitive understanding of the origin of the QCD $\beta-$function
in terms of fundamental interactions in gluon processes and also to
show the equivalence of the $\beta$-function of the coupling constant
of the variational approach to that of QCD.

In order to prove this conjecture, it is necessary to write QCD in the
radiation (also known as Coulomb) gauge.  In this paper a purely gauge
SU(N) Yang-Mills theory in the Hamiltonian formalism is fixed in the
radiation gauge, $\partial_i
A_i^a = 0$.  The time-like and the longitudinal components of the
gauge field are eliminated leaving only transverse gluons and Coulomb
interactions in the action.  Only physical degrees of freedom appear
in the Hamiltonian and there are no ghosts.  The choice of the
radiation gauge, which gives transverse
gluons and Coulomb interactions explicitly, allows direct
comparison with the variational approach.

In section 3, QCD is written in the radiation gauge and the Feynman
rules and diagrams are derived.  In the fourth section, the
modification of the bare Coulomb interaction between two transverse
gluons by the insertion of transverse gluon and transvere gluon -
Coulomb loops, vertex corrections and box diagrams is considered.
This is tantamount to renormalising the coupling constant of the
theory to first order.  The $\beta-$function could be deduced from any
physical process, but to allow direct comparison with the calculation
of the $\beta-$function of the charge of the variational ansatz we consider a
non-local, four transverse gluon vertex with an explicit Coulomb
interaction, present in the infinite series of the action.  In the
next section, though, details of the calculation
of the $\beta-$ function from the variational approach, \cite{8},
necessary for comparison with QCD are highlighted.

\newsection{The $\beta$-function of the Variational Approach.}

From the variational ansatz for the ground state  wavefunctional of an
SU(N) gauge theory the $\beta-$function was deduced in \cite{8}. 
For details the reader is directed to that paper and for more understanding
of the ansatz for the variational approach to the ground state
wavefunctional to \cite{6}.  Here only the necessary details and
results will be highlighted to enable comparison with QCD in the
radiation gauge.

An SU(N) Yang-Mills gauge theory is described by the Hamiltonian,
\begin{equation}
H=\int d^3x\left[\frac{1}{2}E_i^{a2} + \frac{1}{2}B_i^{a2}\right].
\label{v1}
\end{equation}
In the variational calculation of \cite{6} and the analysis of the
ansatz in \cite{8} all terms higher than quadratic in the Hamiltonian
were ignored because they give only small contributions, of $O(g)$, to the
minimization equation.  These terms contained all the information
about gluon-gluon interactions.

The ansatz for the ground state wavefunctional, $\Psi[A]$, was forced
to satisfy the constraint of gauge invariance.  The gauge
transformation $A_i^a \rightarrow A_i^{aU} $ is generated by Gauss'
law, $G^a(x) \Psi[A] = [\partial_i E_i^a(x) - g f^{abc} A_i^b(x)
E_i^c(x)]\Psi[A] = 0$.  Gauss' law gives rise to longitudinal gluons
with a Coulomb interaction.  

In the variational approach the vacuum expectation value of the
Hamiltonian is calculated and minimized.
\begin{equation}
<H> = \frac{1}{Z}\int DA \Psi^* H \Psi
\label{v4}
\end{equation}
Since the Hamiltonian is only considered up to quadratic terms and for
other reasons (calculability, the dominance of a single condensate),
the wavefunctional is written as a Gaussian.  To ensure its invariance
under a gauge transformation it is necessary to sum over the space of
special unitary matrices with the SU(N) group invariant measure.
\begin{equation}
\Psi[A] = \int DU \exp [-\frac{1}{2}\int d^3x d^3y A_i^{aU}(x)
G^{-1}(x-y)A_i^{aU}(y)] 
\label{v5}
\end{equation}
$G(x-y)$ is like a non-local propagator.  $G$ is defined to
coincide with perturbation theory at high momenta and to have a mass
gap.  The value of the mass gap is the variational parameter.
\begin{eqnarray}
G^{-1}(k) = \left\{ \begin{array}{ll} \sqrt{ k^{2} ~} & 
\mbox{ if  $ k^2>M^2$}\\
 M &  \mbox{ if $k^2<M^2$} 
\end{array} 
\right.
\label{v6}
\end{eqnarray}
(\ref{v5}) and (\ref{v6}) together form the variational ansatz.

The calculation of the expectation value of an operator
(e.g. (\ref{v4})) is tantamount to performing a three dimensional path
integral in Euclidean space.  So the exponent of $\Psi^* \Psi$ can in
some sense be considered as an action.  Integration over the field $A$
showed this action to be a non-local, non-polynomial, non-linear sigma
model in three Euclidean dimensions, defined by the action $\Gamma [U]$, where,
\begin{eqnarray}
\Gamma[U] &=& \frac{1}{2} Tr \log {\cal M} +\frac{1}{2} \lambda^a
\Delta^{ac} \lambda^c \nonumber \\
\lambda_i^a &=& \frac{i}{g} tr(\tau^a U^+ \partial_i U)\nonumber \\ 
\Delta^{ac}(x,y) &=& \left[ G(x-y)\delta^{ac} +
S^{ab}(x)G(x-y)S^{Tbc}(y)\right]^{-1} \nonumber \\
S^{ab}(x,y) &=& \frac{1}{2} tr (\tau^a U^+ \tau^b U)
\label{v7}
\end{eqnarray}
$\tau^a$ are generators of an SU(N) group.  Tr is a trace over all
indices, tr is a trace over colour indices and
all summations over indices and integrations over spaces are implicit.
There are no derivatives in ${\cal M}$ - it makes no contribution to
the following analysis and is omitted.

In \cite{8}, this action was analyzed and the $\beta-$function for
the coupling constant of the theory calculated.  The group elements
were decomposed into high and low momentum dependent modes with the
ansatz $U(x) = U_L(x)U_H(x)$.  Writing $U_H(x) = \exp[\frac{ig}{2} \phi^a(x) \tau^a] =
1+\frac{ig}{2}\phi^a \tau^a - \frac{g^2}{8}(\phi^a \tau^a)^2 + O(g^3)$
allowed the effective Lagrangian to be written as,
\begin{eqnarray}
\Gamma_L(x,y) &=& \frac{1}{4}\lambda_{i,L}^a(x) G^{-1}(x-y)
\lambda_{i,L}^a(y)\\ \nonumber
& & + \frac{g}{4}f^{abg}[\lambda_{i,L}^b(x) G^{-1}(x-y)<\phi^g(x)
\partial_i \phi^a(y)> \\ \nonumber
& & + <\partial_i \phi^a(x)\phi^g(y)>  G^{-1}(x-y)
\lambda_{i,L}^b(y)] \\ \nonumber
& & - \frac{g^2}{8}f^{age}f^{bde} \lambda_{i,L}^b(x)G^{-1}(x-y)
\lambda_{i,L}^a(y) [<\phi^g(x)\phi^d(x)> \\ \nonumber
& & +<\phi^g(y)\phi^d(y)> -
2<\phi^g(x)\phi^d(y)>] \\ \nonumber
& & + \frac{g^2}{16} \lambda_{i,L}^a(x) G^{-1}(x-y) \lambda_{i,L}^a(y)
\frac{C_2(G)}{tr[\delta^{aa}]} [<\phi^b(x)\phi^b(x)> \\ \nonumber
& & +<\phi^b(y)\phi^b(y)>
-2<\phi^b(x)\phi^b(y)>].
\label{v9}
\end{eqnarray}
This is an effective low energy Lagrangian.  High momentum modes in
the region $M'<k<\Lambda$ were integrated over, where $\Lambda$ is the
UV cut-off.  The scale $M'$ is arbitrary but in \cite{8} and here
$M'=M$ is chosen for simplicity of calculation and clarity of
presentation.  Hence, $\phi^a$ depends only upon high momentum modes
and $\lambda_{i,L}^a$ depends only upon low momentum modes.

Now the correlations of $\phi^a$ need to be calculated.
\begin{equation}
<\phi^a(x)> = 0
\label{v10}
\end{equation}
\begin{equation}
<\phi^a(x) \phi^b(x)> = \frac{\delta^{ab}}{\pi^2} \log[\frac{\Lambda}{M}]
\label{v11}
\end{equation}
\begin{equation}
<\phi^a(x) \phi^b(y)> = \left\{ \begin{array}{cc} C & \mbox{ $M|x-y|>\mu$} \\
-\frac{\delta^{ab}}{\pi^2} \log [M|x-y|] +K & \mbox{ $M|x-y|<\mu$} 
\end{array}
\right.
\label{v12}
\end{equation}
$C$ and $K$ are finite contributions which are independent of $M$ and
$\Lambda$ and so are ignored in the following.  (\ref{v11}) occurs in $\Gamma_L$ in terms such as,
\begin{equation}
\frac{g^2}{16} \lambda_{i,L}^a(x) G^{-1}(x-y) \lambda_{i,L}^a(y)
\frac{C_2(G)}{tr[\delta^{aa}]} <\phi^b(x) \phi^b(x)>.
\label{v13}
\end{equation}
\begin{figure}
\epsfbox{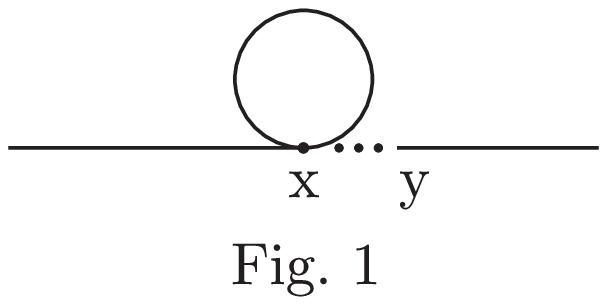}
\label{f2}
\end{figure}
This can be represented by a Feynman diagram as in Fig. 1, which shows
a tadpole diagram with a non-local connection.  The external lines
correspond to low momentum fields and the internal solid loop
represents the integration over high momentum fields.  The dotted line
corresponds to $G^{-1}(x-y)$ which contains information about
transverse gluons.  In the region $k<M$, $G^{-1}(x-y) = M\delta(x-y)$
and the propagator becomes local, associating the ends of the dotted
line, and the standard tadpole diagram is recovered.

(\ref{v12}) appears in $\Gamma_L$ in terms such as,
\begin{equation}
-\frac{g^2}{8} \lambda_{i,L}^a(x) G^{-1}(x-y)\lambda_{i,L}^a(y)
\frac{C_2(G)}{tr[\delta^{aa}]}<\phi^b(x)\phi^b(y)>,
\label{v14}
\end{equation}
which can be interpreted in terms of `horse-shoe' diagram, Fig. 2.
The external, internal and dotted lines correspond to the same
quantities as in Fig. 1, described above.  This diagram is unimportant
in the region $\frac{\mu}{M}<|x-y|$ as then it is finite.  The
interpretation of the appearance of $|x-y|$ must come from the
Lagrangian.
\begin{figure}
\epsfbox{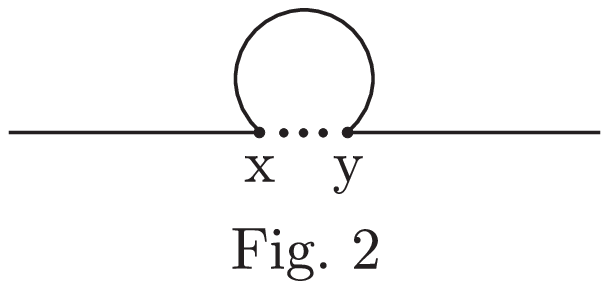}
\label{f1}
\end{figure}

The effective Lagrangian can now be written as,
\begin{equation} 
\Gamma_L(x,y) = \frac{1}{4} \tilde{\lambda}_{i,L}^a(x) G^{-1}(x-y)
\tilde{\lambda}_{i,L}^a(y),
\label{v15}
\end{equation}
where, 
\begin{eqnarray}
\tilde{\lambda}_{i,L}^a &=& \frac{i}{\tilde{g}}{\rm tr}[\tau^a U_L^+ \partial_i
U_L] \nonumber \\
\tilde{g} &=& g+\frac{g^3}{(4\pi)^2}4C_2(G)\log\left[\frac{\Lambda}{\tilde{M}}\right]
+O(g^5) 
\label{v16}
\end{eqnarray}
and,
\begin{eqnarray}
\tilde{M} = \left\{ \begin{array}{cc} M & \mbox{ $|x-y|>\frac{\mu}{M}$} \\
\frac{1}{|x-y|} & \mbox{ $|x-y|<\frac{\mu}{M}$}.
\end{array}
\right.
\label{v17}
\end{eqnarray}
From this the $\beta$-function is easily calculated.
\begin{equation}
\beta(g) = M'\frac{\partial}{\partial M'}\tilde{g}|_{g,\Lambda} =
-\frac{g^3}{(4\pi)^2} 4 C_2(G) +O(g^5).
\label{v18}
\end{equation}
This result is close to the QCD $\beta$-function calculated in
\cite{1}, \cite{2}, \cite{3} and \cite{4} as discussed in \cite{8}.
This prompted the conjecture that the anti-screening contribution of
`$-4$' to the QCD $\beta$-function is due to the renormalisation of a
Coulomb interaction.  Such Coulomb interactions originate with Gauss'
law which can be used as the generators of SU(N) gauge transformation,
i.e. as the generators of U.  Further, the conjecture was made that
the `$\frac{1}{3}$' screening contribution must be associated with
transverse gluon processes - exactly the interaction terms omitted
from the Hamiltonian in the the exploratory variational calculations
of \cite{6} as they only give small, $O(g)$, contributions to the
minimization equation.  This conjecture is reliant upon the validity
of the original ansatz for the ground state wavefunctional. 
In the next two sections QCD is written in the radiation gauge and
this conjecture is proved.  Thus the equivalence of the
$\beta$-function of the coupling constant written in the variational
ansatz to that of QCD is shown and a physical understanding of the
origins of screening and anti-screening effects in gluodynamics is obtained.

\newsection{QCD in the radiation gauge.}

SU(N) Yang-Mills gauge field theory in the Hamiltonian formalism is
defined in (\ref{v1}) with the constraint of Gauss'
law, $D_i \pi_i^a = 0$.  The covariant derivative is defined as,
\begin{equation}
D_{\mu} w^a = \partial_{\mu} w^a -gf^{abc}A_{\mu}^b w^c .
\label{1.5}
\end{equation}
The generators of the SU(N) internal symmetry, $T^a$, and the group
structure constants, $f^{abc}$, have the following properties:
\begin{eqnarray}
[T^a,T^b] &=& if^{abc}T^c \nonumber \\
Tr[T^a T^b] &=& \frac{1}{2} \delta^{ab} \nonumber \\
f^{abc}f^{abd} &=& C_2(G)\delta^{cd}.
\label{2}
\end{eqnarray}
$C_2(G)$ is the second Casimir operator and $G$ denotes the adjoint
representation in this case.  For SU(N), $C_2(G)=N$.  Calculations
will be performed in
Euclidean space.

Imposing the radiation gauge, $\partial_i A_i^a = 0$, allows us to
write the Hamiltonian density,
\begin{eqnarray}
H &=& \frac{1}{2}\pi_i^a \pi_i^a
+\frac{1}{2}\partial_iA_j^a\partial_iA_j^a
-gf^{abc}\partial_iA_j^aA_i^bA_j^c \nonumber \\
& & +\frac{g^2}{4}f^{abc}f^{ade}A_i^bA_j^cA_i^dA_j^e .
\label{5}
\end{eqnarray}

From the equivalent to this Hamiltonian for the SU(2) Yang-Mills
theory in Minkowski space Khriplovich \cite{1} deduced the Feynman diagrams
and wrote the full Green's functions of the theory using the spectral
representation in the radiation gauge.  This process
involves decomposing
the canonical momenta into transverse and longitudinal components and
solving to eliminate the longitudinal component.  The time-like
component of the vector field is also eliminated and the Coulomb
interaction becomes explicit.  
The final Hamiltonian obtained, with the canonical momenta solved in
terms of the components of the vector field, is an infinite series in
$g$ which contains only transverse gluons and a Coulomb interaction. 
Khriplovich's interpretation was that such an infinite series in
the coupling constant of the theory was the price for having no
unphysical ghost fields.  A similar procedure will be used for
the more complex case of the SU(N) Yang-Mills theory in Euclidean
space.  A
diagrammatic treatment of the theory will be used which will provide physical
insight into the processes from which the screening and anti-screening
contributions to the $\beta$-function originate, in contrast to the
spectral representation used in \cite{1}.  The Hamiltonian for a purely SU(N) gauge theory
shall be written containing only transverse gluons and
Coulomb interactions.  This allows a direct comparison to be made
between the variational ansatz for the groundstate wavefunctional and QCD. 

To start, the canonical momentum is decomposed into longitudinal and
transverse parts, $p_i^a$ and $\partial_i \phi^a$ respectively,
\begin{eqnarray}
\pi_i^a &=& p_i^a+\partial_i \phi^a \nonumber \\
\partial_i p_i^a &=& 0.
\label{8}
\end{eqnarray}
Thus, we can write the Hamiltonian density,
\begin{eqnarray}
H &=& \frac{1}{2}p_i^ap_i^a+\frac{1}{2} \partial_iA_j^a\partial_i
A_j^a -gf^{abc}\partial_iA_j^a A_i^b A_j^c \nonumber \\
& & +\frac{g^2}{4}f^{abc}f^{ade}A_i^bA_j^c A_i^d A_j^e -
\frac{1}{2}\phi^a \Delta \phi^a.
\label{9}
\end{eqnarray}

Substituting this decomposition of the canonical momenta into Gauss'
law gives the relation,
\begin{equation}
\Delta \phi^a = gf^{abc}A_i^b(p_i^c+\partial_i\phi^c).
\label{10}
\end{equation}
This expression can be solved for $\phi^a$ by a process of
reiteration with the definition of the inverse Laplacian, 
\begin{eqnarray}
\Delta^{-1}(x) f(x) &=& \int d^4x \Delta^{-1}(x-x')f(x') \nonumber \\
\Delta (x) \Delta^{-1} (x-x') &=& \delta^4(x-x').
\label{delt}
\end{eqnarray}
$\phi^a$ therefore becomes an infinite series in $g$,
\begin{eqnarray}
\phi^a(x) &=& \phi_1^a(x) +\phi_2^a(x)+\phi_3^a(x)+O(g^4) \nonumber \\
\phi_1^a(x) &=& gf^{abc}\int d^4x'
\Delta^{-1}(x-x')[A_i^b(x')p_i^c(x')] \nonumber \\
\phi_2^b(x) &=& g^2f^{abc}f^{cde}\int d^4x' \int d^4x''
\Delta^{-1}(x-x') [A_i^b(x') \partial_i^{x'} \Delta^{-1}(x'-x'')
[A_j^d(x'') p_j^e(x'')]] \nonumber \\
\phi_3^a(x) &=& g^3f^{abc}f^{cde}f^{efg} \int d^4x' \int d^4x'' \int d^4x'''
\Delta^{-1}(x-x') [A_i^b(x')\partial_i^{x'} \Delta^{-1}(x'-x'')
\nonumber \\
& & [A_j^d(x'') \partial_j^{x''} \Delta^{-1}(x''-x''')
[A_k^f(x''')p_k^g(x''')]]]
\label{12}
\end{eqnarray}
where $\partial_i^{x'}=\frac{\partial}{\partial x_i'}$.

By substituting (\ref{12}) into (\ref{9}) and identifying $p_i^a$ with
$-\partial_0 A_i^a$, we obtain the following
Hamiltonian,
\begin{eqnarray}
H &=& \int d^4x[-\frac{1}{2}A_i^a \Box A_i^a -gf^{abc}\partial_i A_j^a
A_i^b A_j^c
+\frac{g^2}{4} f^{abc}f^{ade}A_i^bA_j^cA_i^dA_j^e] \nonumber \\
& & -\frac{g^2}{2}f^{abc}f^{ade}\int d^4x d^4x'[A_i^b(x)\partial_0 A_i^c(x)]
\Delta^{-1}(x-x') [A_j^d(x')\partial_0 A_j^e(x')] \nonumber \\
& & -g^3f^{abc} f^{cde} f^{afg} \int d^4x d^4x' d^4x'' [A_i^b(x)
\partial_i^x \Delta^{-1}(x-x')[A_j^d(x') \partial_0 A_j^e(x')]]
\Delta^{-1}(x-x'') \nonumber \\
& &\;\;\;\;\;\; [A_k^f(x'') \partial_0 A_k^g(x'')] \nonumber \\
& & -\frac{3}{2}g^4 f^{abc} f^{cde} f^{afg} f^{ghi} \int d^4x d^4x'
d^4x'' d^4x''' [A_i^b(x) \partial_i^x \Delta^{-1}(x-x') [A_j^d(x')
\partial_0 A_j^e(x')]] \nonumber \\
& & \;\;\;\;\;\;\Delta^{-1}(x-x'') [A_k^f(x'') \partial_k^{x''} \Delta^{-1}(x''-x''')
[A_l^h(x''') \partial_0 A_l^i(x''')]] \nonumber \\
& & +O(g^5)
\label{13}
\end{eqnarray}

This Hamiltonian contains three terms familiar from the usual
Lorentz covariant treatment of QCD and an infinite series of non-local
terms, the non-locality provided by the definition of $\Delta^{-1}$.
The transverse gluon propagator, denoted by the conventional
`pig-tails', is given by,
\begin{eqnarray}
D_{ij}^{ab(0)}(p) &=& \delta^{ab}\frac{1}{(p^2+p_0^2)} d_{ij}(p)
\nonumber \\
d_{ij}(p) &=& \delta_{ij}-\frac{p_ip_j}{p^2}
\label{14}
\end{eqnarray}
where $p^2 = p_i^2$.  The superscript of zero denotes that this is the bare
propagator.  The second and third terms give rise to the usual three- and four-
gluon interaction vertices but in this Hamiltonian these are
interactions of transverse gluons only.  The three transverse gluon
interaction with its Feynman rule is shown in Fig. 3.  The convention that positive
momentum flows into a vertex is used and the sum of momenta into a
vertex is zero, (i.e. in Fig. 3, $p+q+r=0$).

\begin{figure}
\epsfbox{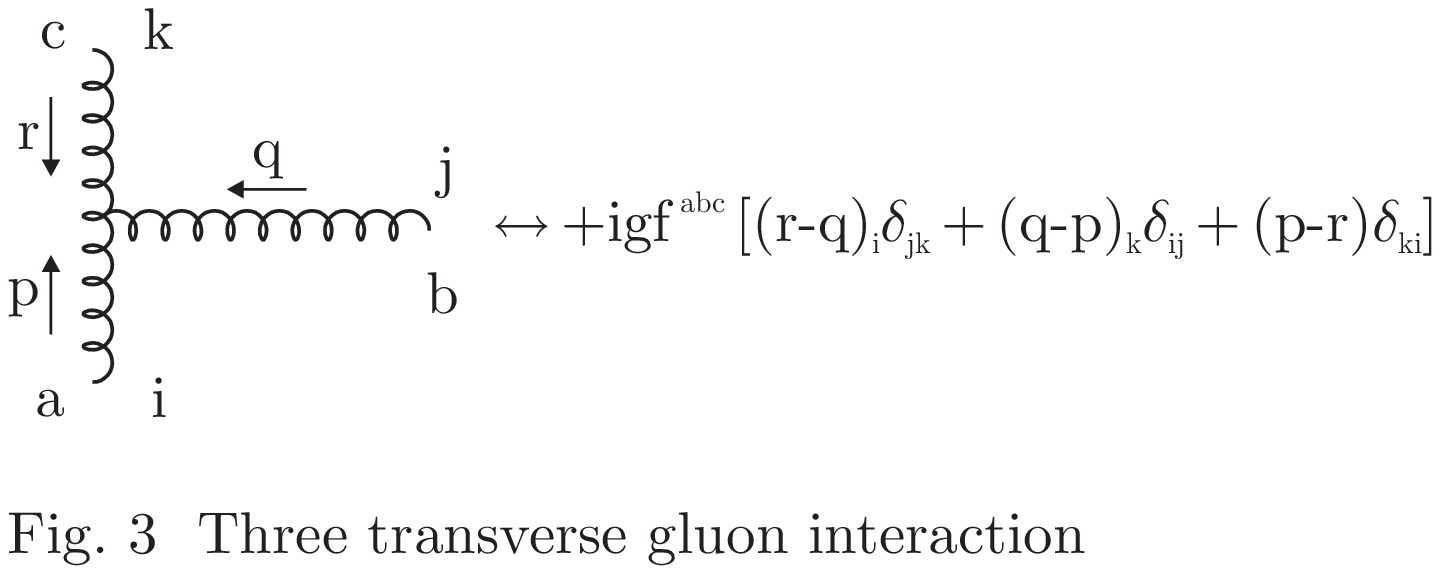}
\end{figure}

The fourth, fifth, sixth and all subsequent terms in the action form
an infinite series in the coupling constant, $g$.  Each term involves
a Coulomb interaction between two gluons and each successive term
differs from the previous by the emission of a transverse gluon from
the Coulomb interaction.  The Feynman diagrams and fully symmetrized
rules corresponding to the fourth, fifth and sixth terms in the action
(\ref{13}) are given in Figs 4, 5 and 6, respectively.

\begin{figure}
\epsfbox{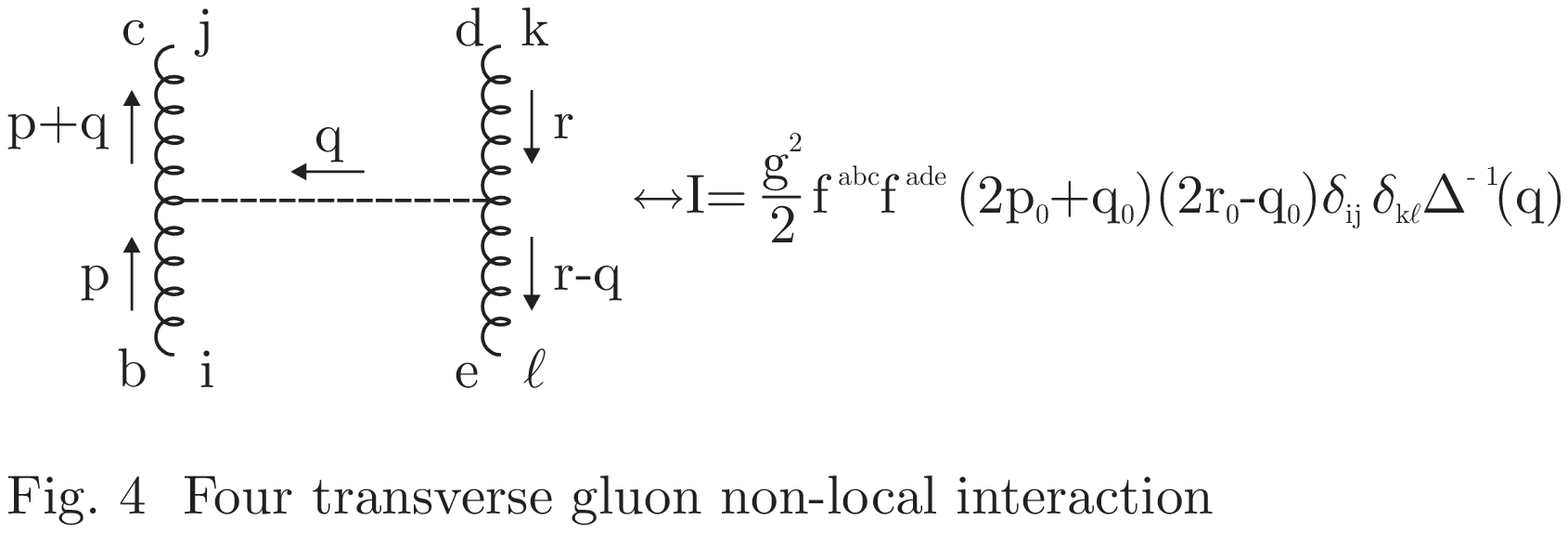}
\end{figure}
\begin{figure}
\epsfbox{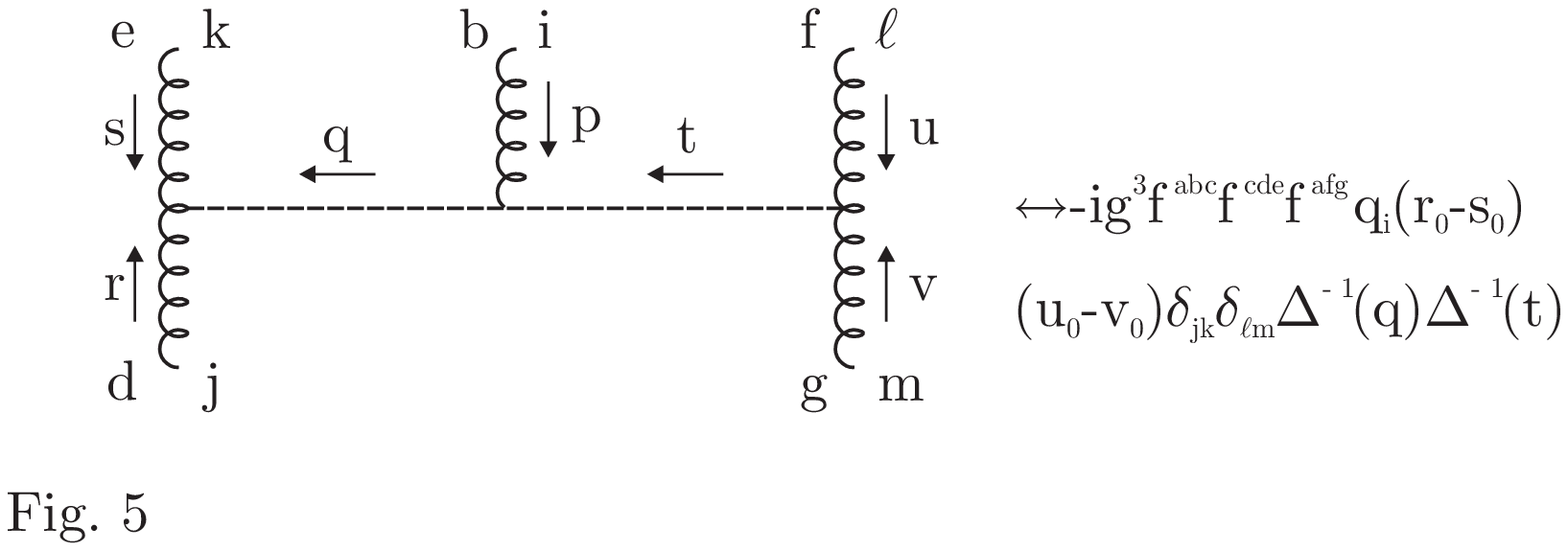}
\end{figure}
\begin{figure}
\epsfbox{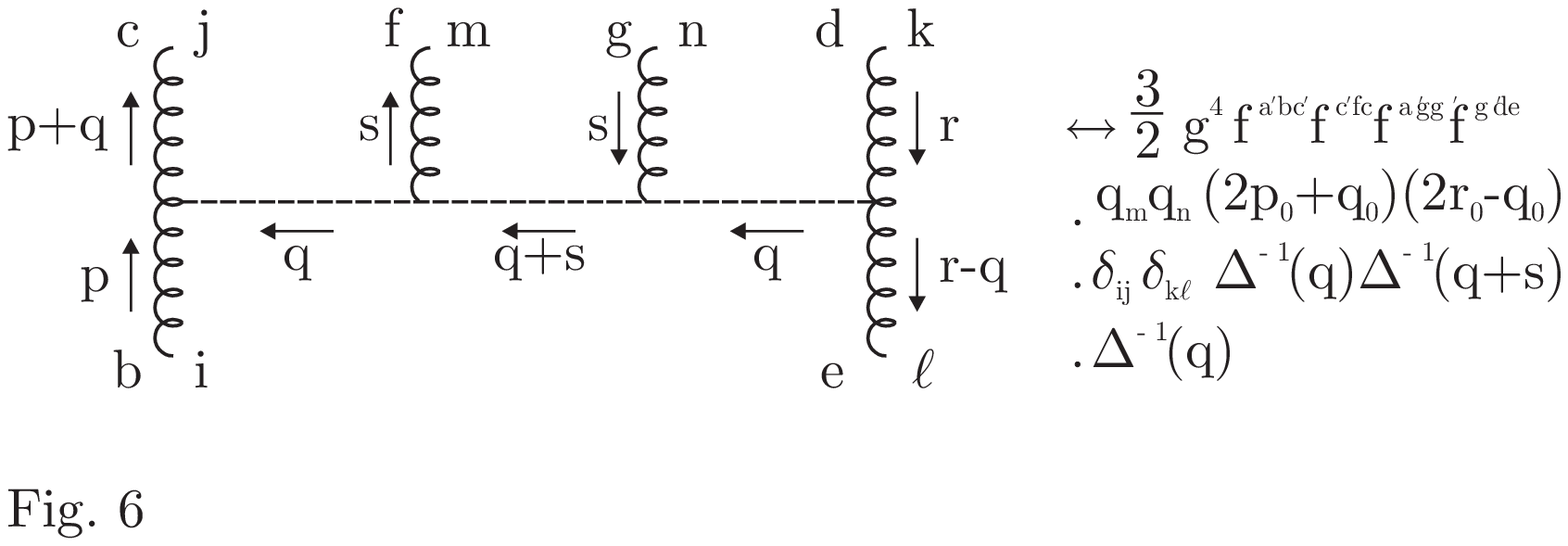}
\end{figure}

The Coulomb interaction is denoted by a dashed line and is written as,
\begin{equation}
D_{00}^{ab(0)}(p) = -\frac{\delta^{ab}}{p^2},
\label{15}
\end{equation}
where, again, $p^2=p_i^2$ and the zero superfix denotes a bare
interaction.  It is first order modifications to the bare Coulomb interaction
between two transverse gluons, Fig. 4, that are considered in the next
section and it is from this that the origins of screening and
anti-screening effects in gluon processes are deduced.

\newsection{Renormalisation of the Coulomb Interaction to First Order}

It is possible to deduce the $\beta$-function from any physical
process.  The case that shall be studied in the following can be
regarded as the renormalisation of a four transverse gluon
non-local interaction.  Two transverse gluons meet at a point, $x$
say, and two more at another point, $x'$, with an interaction between
the vertices at $x$ and $x'$.  To allow a fruitful comparison with the
calculation of the $\beta$-function from the variational approach a Coulomb
interaction between points $x$ and $x'$ is considered.  This is
distinguishable from the case with a transverse gluon exchanged between
$x$ and $x'$ by the poles of the propagator.  Such a four transverse
gluon non-local vertex can also be considered as a Coulomb interaction
between two transverse gluons taking the role of external
charges.  Of course, more
complicated interactions involving more Coulomb and transverse gluon
propagators and vertices are possible.  These modifications of the
bare propagators and vertices are constructed by combining terms of
lower order in $g$ (e.g. combining three and four transverse gluon
vertices with Fig.'s 4 and 5) or from closing loops
in terms of higher order in g (e.g. Fig. 6).  The coupling constant
can then be redefined to absorb the infinities associated with these
modifications and from this the $\beta$-function deduced.  In this
section, all diagrams that contribute to the $\beta$-function up to
first order will be calculated.

From previous calculations in QCD, it is known that box diagrams do not
contribute to the $\beta$-function.  Choosing the radiation gauge maps
these gluon box diagrams into a number of diagrams formed of
combinations of transverse gluon and Coulomb interactions.  Each of
these new box diagrams may be individually divergent but summed
together they cannot contribute to the QCD $\beta$-function.

The other diagrams that need be considered for the renormalisation of
a Coulomb interaction to first order are loops in the Coulomb
interaction, vertex corrections and tadpole contributions.  Transverse
gluon and transverse gluon - Coulomb loops are both divergent.  A
purely Coulomb loop is not possible by the Feynman rules of the
previous section.  One of the two possible vertex corrections is zero
and the other contributes an infinity.  These cases will be considered
in the following subsections.  

First, it is necessary to see that
there is no contribution from the tadpole correction to the Coulomb
interaction, Fig. 7, which is formed from the combination of Fig.'s 3
and 5.  With two ends of Fig. 3 closed to form a loop, the product of
a delta function in the colour indices and the totally anti-symmetric
group structure constant gives identically zero.

\begin{figure}
\epsfbox{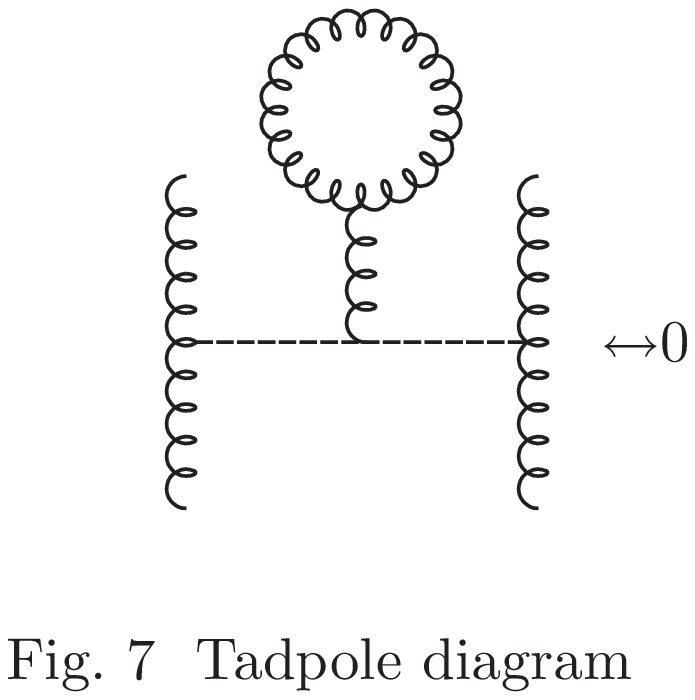}
\end{figure}

\newsubsection{Renormalisation of a Coulomb Interaction by a
Transverse Gluon - Coulomb Loop.}

The Coulomb interaction can be modified by considering the insertion
of a transverse gluon - Coulomb loop.  This is equivalent to the
emission and reabsorbance of a transverse gluon by the Coulomb
interaction.  This is exactly the diagram formed by joining the two
`internal' transverse gluon lines in Fig. 6.  

\begin{figure}
\epsfbox{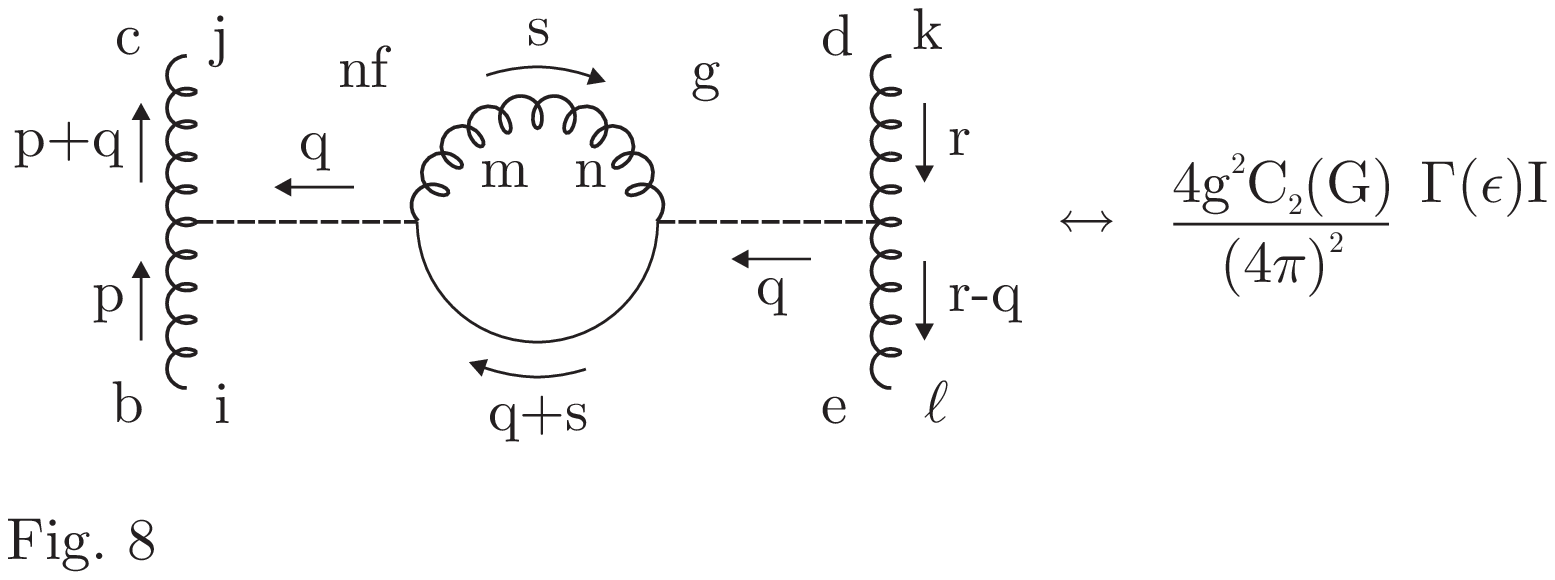}
\end{figure}

Fig. 4 shows the four transverse gluon non-local vertex with a Coulomb
interaction.  This diagram is denoted by I.  Fig. 8 shows a Coulomb
interaction modified by a transverse gluon - Coulomb loop between two
external transverse gluons.  This interaction is written as,
\begin{eqnarray}
& & \frac{3}{2}g^4 f^{afh} f^{hbc} f^{af'g} f^{gde} \int
\frac{d^4s}{(2\pi)^4} q_m q_n (2p_0+q_0) (2r_0-q_0) \delta_{ij}
\delta_{kl} \Delta^{-1}(q) \Delta^{-1}(q+s) \nonumber \\
& & \Delta^{-1}(q)
\frac{\delta^{ff'}d_{mn}(s)}{[s^2+s_0^2]} = {\rm I} \Sigma,
\label{r1}
\end{eqnarray}
where,
\begin{equation}
\Sigma = -3g^2 C_2(G) q_m q_n \Delta^{-1}(q) \int \frac{d^4s}{(2\pi)^4}
\frac{(\delta_{mn} s^2 -s_m s_n)}{s^2(s^2+s_0^2)(s^2+2sq+q^2)}.
\label{r2}
\end{equation}
It is important to note that since Lorentz invariance has been broken
by the choice of gauge the time-like component is not treated the same
as the spatial components of momentum in the loop integration.
$\Sigma$ is evaluated in Appendix B with the result,
\begin{equation}
\Sigma=\frac{4g^2 C_2(G)}{(4\pi)^2}\Gamma(\epsilon),
\label{x1}
\end{equation}
where the constants are the same as in (\ref{v18}), $\Gamma(\epsilon)$
is defined
in the standard manner and given in Appendix A and $\epsilon = w - A$
from the integration with the dimensional regularisation
formulae, also given in Appendix A.  For QCD in $3+1$ dimensions $\epsilon \rightarrow 0$.  All
finite terms are ignored in the calculation of the $\beta$-function.

\newsubsection{Renormalisation of a Coulomb Interaction by a Transverse
Gluon Loop.}

From the combination of two four transverse gluon, non-local vertices,
Fig. 4, a box diagram and a Coulomb interaction with a transverse
gluon loop can be made.  The fact that box diagrams do not contribute
to the $\beta$-function has already been discussed above.  Fig. 9
shows a Coulomb interaction with a transverse gluon loop inserted.
This interaction is written as,
\begin{eqnarray}
& & g^4f^{abc} f^{ad'e'} f^{a'b'c'} f^{a'de} \int
\frac{d^4u}{(2\pi)^4} (2p_0+r_0) (2u_0+r_0)^2 (2s_0-r_0) \delta_{ij}
\delta_{k'l'} \delta_{i'j'} \delta_{kl} \Delta^{-1}(r) \Delta^{-1}(r)
\nonumber \\
& & \frac{\delta^{d'c'} d_{k'j'}(u)}{[u^2+u_0^2]} \frac{\delta^{e'b'}
d_{i'l'}(r+u)}{[(r+u)^2 + (r_0+u_0)^2]} = {\rm I}\Pi,
\label{r6}
\end{eqnarray}
where,
\begin{equation}
\Pi = -\frac{g^2}{2} C_2(G) \int \frac{d^4u}{(2\pi)^4} (2u_0+r_0)^2
\Delta^{-1}(r) ND.
\label{r7}
\end{equation}

\begin{figure}
\epsfbox{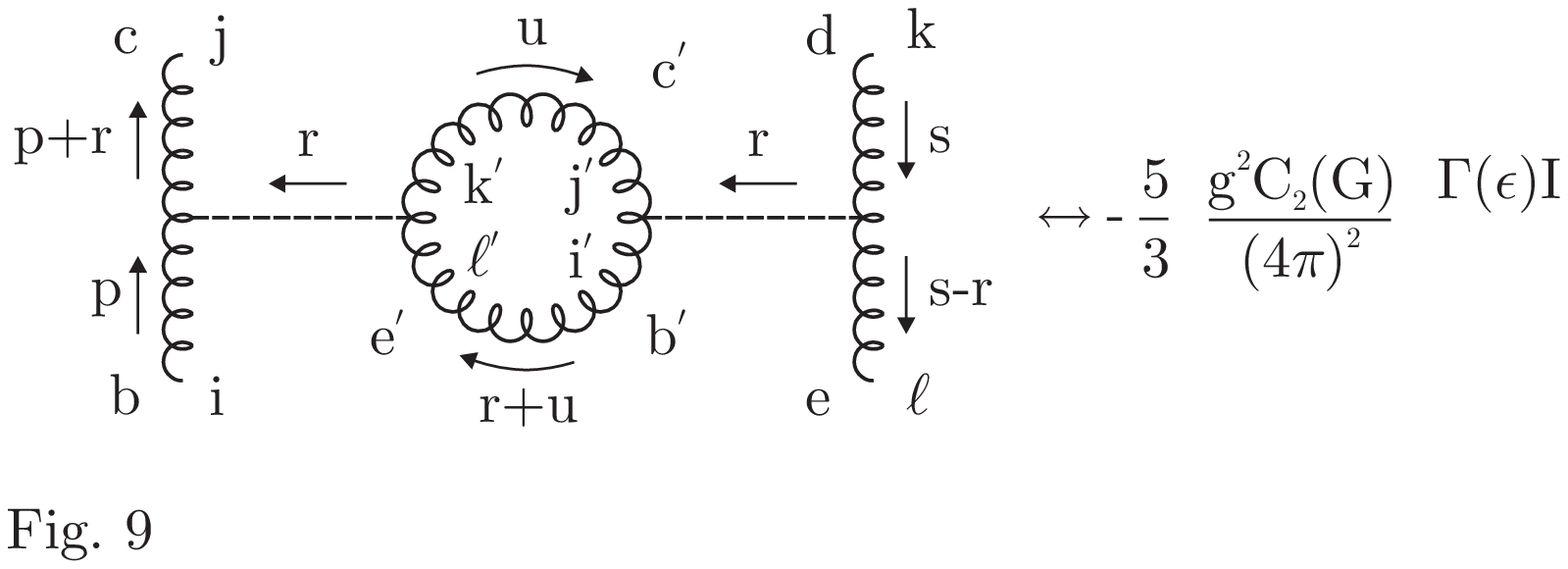}

\end{figure}

The numerator and denominator, $N$ and $D$, are,
\begin{eqnarray}
N &=& \delta_{i'j'} \delta_{k'l'} (u^2\delta_{k'j'}-u_{k'}
u_{j'})((r+u)^2 \delta_{i'l'} - (r+u)_{i'} (r+u)_{l'}) \nonumber \\
D &=& \left[ u^2 (u^2+u_0^2) (r^2 +2ru +u^2) (r^2+ 2ru +u^2 +r_0^2
+2r_0u_0 +u_0^2)\right]^{-1}
\label{r8}
\end{eqnarray}
$\Pi$ is evaluated by the lengthy calculations of Appendix C.
Throughout the calculation it was unobvious that the time-like
components of momentum would cancel but there it is seen that they do.
We finally write the modification to the Coulomb interaction by the
insertion of a transverse gluon loop as,
\begin{equation}
\Pi = -\frac{5}{3} \frac{g^2 C_2(G)}{(4\pi)^2} \Gamma(\epsilon)
\label{r17}
\end{equation}

\newsubsection{Calculation of Vertex Corrections}

As well as modifications to the Coulomb interaction by the insertion
of loops, corrections to the gluon - gluon - Coulomb vertex are
possible.  To first order, two such corrections are possible.

\begin{figure}
\epsfbox{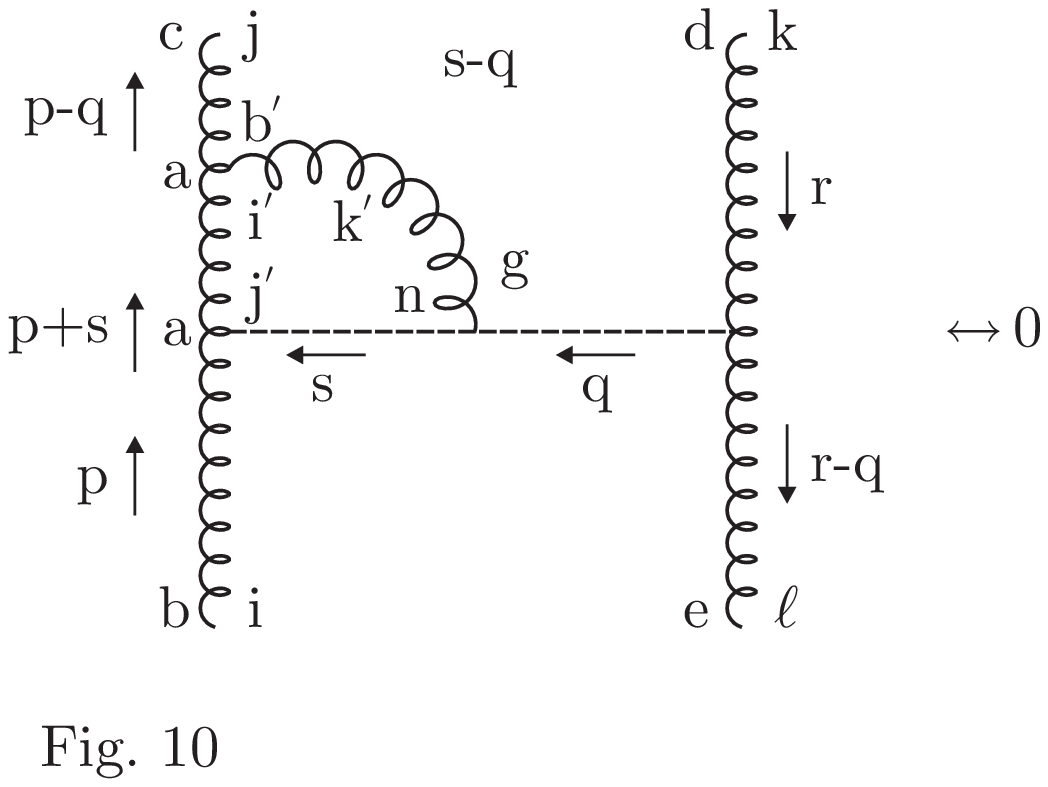}
\end{figure}
\begin{figure}
\epsfbox{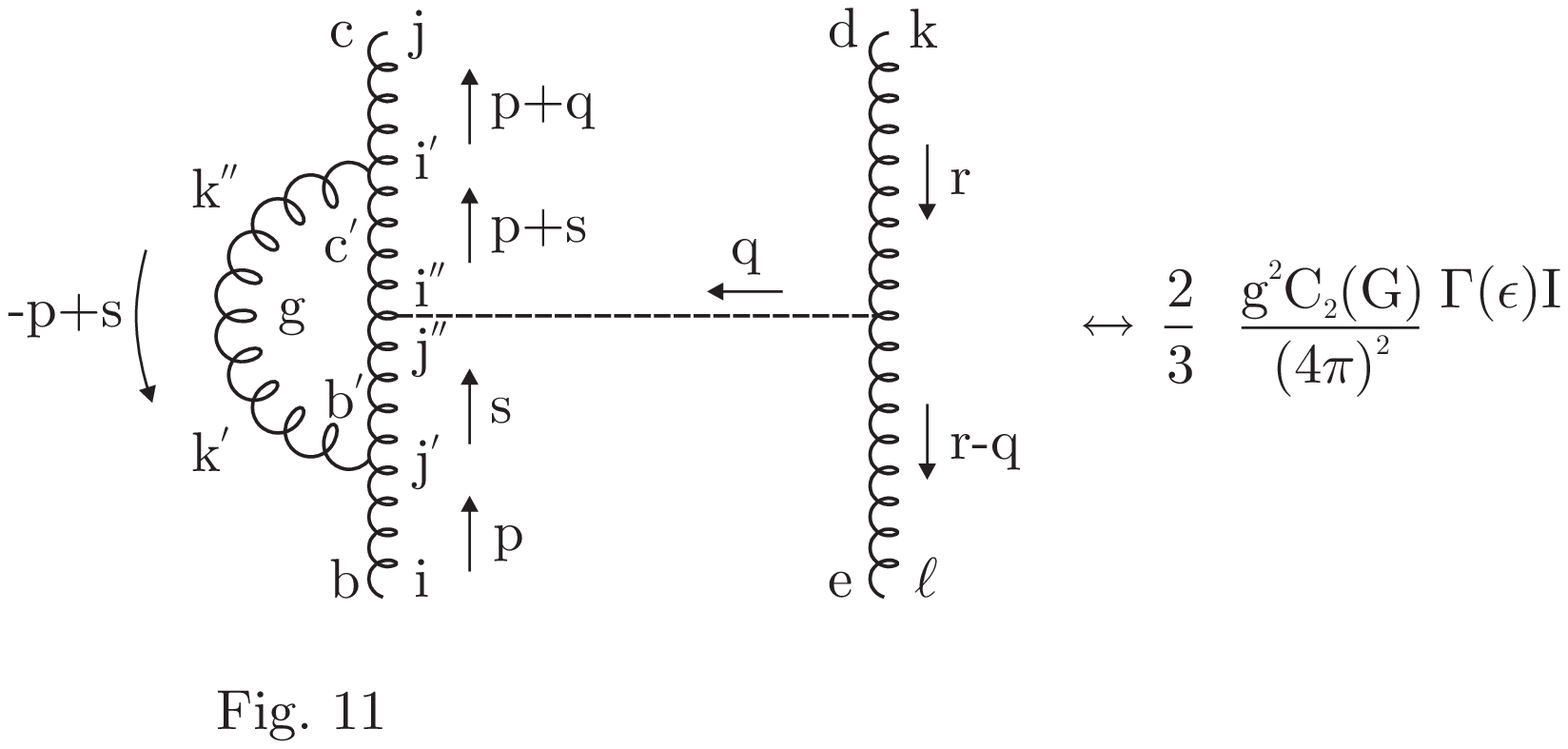}

\end{figure}

The first correction is obtained by combining the interactions of
Fig. 3 and 5 and is shown in Fig. 10.  This is written as,
\begin{equation}
-\frac{g^4}{2} f^{agf} f^{fbh} f^{ade} f^{hgc} \delta_{kl} (2r_0-q_0)
 \Delta^{-1}(q) \int \frac{d^4 s}{(2\pi)^4} (2p_0+s_0) ND,
\label{r18}
\end{equation}
where,
\begin{eqnarray}
N &=& (-s+q)_m\delta_{ij'} [(s-p)_{i'}\delta_{k'j}
+(p-q)_j\delta_{i'k'} + (2p+s-q)_{k'}\delta_{i'j}] \nonumber \\
& & ((p+s)^2\delta_{j'i'} -(p+s)_{j'} (p+s)_{i'})((s-q)^2 \delta_{k'm}
- (s-q)_{k'}(s-q)_{m}), \nonumber \\
D &=& [s^2 (p^2 +2ps +s^2) (s^2 +2ps +p^2 +s_0^2 +2p_0 s_0 + s_0^2)
(s^2 -2sq +q^2) \nonumber \\
& & (s^2 -2sq +q^2 + s_0^2 - 2s_0 q_0 +q_0^2)]^{-1}.
\label{r19}
\end{eqnarray}
This expression is evaluated in Appendix D.  A numerator of $O(s^6)$,
where $s$ is the momentum variable around the loop, is required for
this diagram to be divergent.  Although this criterion initially seems
to be satisfied, all terms of such order cancel exactly.  Hence,
Fig.10 is convergent and makes no contribution to the QCD
$\beta$-function.

The second possible vertex correction is made by the combination of
two triple transverse gluon vertices, Fig. 3, and a four transverse
gluon, non-local vertex, Fig. 4.  From these components a box diagram
is also possible but such diagrams have already been discussed above.
The second vertex correction is shown in Fig. 11 and written as,
\begin{equation}
-\frac{g^4}{2} f^{ab'c'} f^{ade} f^{c'cg} f^{bb'g} \int \frac{d^4
s}{(2\pi)^4} (2s_0 +q_0)(2r_0-q_0) \delta_{kl} \Delta^{-1}(q) ND,
\label{r22}
\end{equation}
where,
\begin{eqnarray}
N &=& \delta_{j''i''} [(2p-s+q)_{i'}\delta_{jk''} + (-p-2q-s)_{k''}
\delta_{i'j} + (2s+q-p)_j \delta_{i'k''}] \nonumber \\
& & [(2s-p)_i \delta_{j'k'}
+(-s-p)_{k'} \delta_{ij'} + (2p-s)_{j'}\delta_{ik'}] \nonumber \\
& & [(-p+s)^2 \delta_{k''k'}
-(-p+s)_{k''}(-p+s)_{k'}] [s^2\delta_{j'j''} -s_{j'}s_{j''}] \nonumber
\\
& &[(q+s)^2
\delta_{i''i'} -(q+s)_{i''}(q+s)_{i'}], \nonumber \\
D &=& [s^2 (s^2+s_0^2) (p^2 -2ps +s^2) (p^2-2ps+s^2 +p_0^2 -2p_0s_0
+s_0^2) \nonumber \\
& & (q^2+2qs +s^2)(q^2+2qs+s^2 +q_0^2 +2q_0s_0 +s_0^2)]^{-1}.
\label{r23}
\end{eqnarray}
This  expression is evaluated in Appendix E, where the Coulomb
interaction between two transverse gluons with a
transverse gluon correction about one of the gluon - gluon - Coulomb
vertices is finally written as I$\Gamma$, with,
\begin{equation}
\Gamma = \frac{2}{3} \frac{g^2 C_2(G)}{(4\pi)^2}\Gamma(\epsilon).
\label{r32}
\end{equation}
Two such vertex corrections are possible to first order, one at either
end of the
Coulomb interaction.  So, the modification to the Coulomb
interaction between
a pair of external transverse gluons to second order in $g$ is,
\begin{equation}
I(\Sigma +\Pi +2\Gamma) = I\left[\frac{g^2
C_2(G)}{(4\pi)^2}\Gamma(\epsilon)(4-\frac{1}{3})\right].
\label{r33}
\end{equation}

With some normalisation, a redefinition of the coupling
constant written in I is seen to lead exactly to the QCD
$\beta$-function.  The anti-screening contribution of `$-4$' is seen
to come from the
renormalisation of a Coulomb interaction by a transverse gluon.  This
is exactly the sum of Fig.'s 1 and 2 from the variational approach,
where $G^{-1}$ contains information about transverse gluons.  The
`$\frac{1}{3}$' contribution is seen to arise from a screening
contribution of `$\frac{5}{3}$' from a transverse gluon loop and an
anti-screening contribution of `$-\frac{4}{3}$' from two vertex
corrections.  These contributions come from the interaction of gluons,
exactly the terms omitted from the exploratory variational calculation
of \cite{6}.  

\newsection{Conclusion}

In the exploratory variational calculation of \cite{6}, the assumption was made
that the coupling constant written in the variational ansatz for the
ground state wavefunctional of QCD ran as the coupling constant of
QCD.  This assumption was found to be nearly correct in \cite{8} where
the $\beta$-function of coupling constant of the variational ansatz
was calculated. It was found to have only a `$-4$' anti-screening
contribution and not both the known `$-(4-\frac{1}{3})$' screening and
anti-screening contributions of QCD.  The only physical constraint
imposed upon the variational ansatz was Gauss' law.  The
$\beta$-function from the variational approach was calculated from a
non-linear, non-local sigma model in three Euclidean dimensions
written in terms of elements of the gauge group for which Gauss' law
can be used as the generators.  Gauss' law gives rise to longitudinal
gluons with Coulomb interaction.  Hence, the conjecture was made in
\cite{8} that the anti-screening `$-4$' contribution must come from
the renormalisation of a Coulomb interaction and the screening
`$\frac{1}{3}$' contribution must arise from gluon interactions -
exactly the non-linear terms missed out of the original calculations, \cite{6}.

This conjecture, however, rested heavily upon the validity of the
original ansatz.  In order to prove this conjecture to be correct, it
was explicitly checked in this paper by performing calculations for
QCD in the Hamiltomian formalism and choosing the radiation (or
Coulomb) gauge.  The verification of the conjecture suggests that the
original ansatz captures the anti-screening behaviour of
gluons in QCD that leads to asymptotic freedom and is essential to
address strong coupling problems.  Further, a full
understanding of the origins of screening and anti-screening effects
in terms of gluon interactions has been obtained.

The anti-screening contribution of `$-4$' is seen
to come from the
renormalisation of a Coulomb interaction by a transverse gluon.  This
is shown to be exactly equivalent to the anti-screening contribution
obtained from the variational approach.  The
`$\frac{1}{3}$' contribution is seen to arise from a screening
contribution of `$\frac{5}{3}$' from a transverse gluon loop and an
anti-screening contribution of `$-\frac{4}{3}$' from two vertex
corrections.  These contributions come from the interaction of gluons,
exactly the terms omitted from the exploratory variational calculation
of \cite{6}.  It is hoped, that by a more refined consideration of the
initial ansatz to include the non-linear terms that correspond to
gluon interactions, it can be written to capture the exact running of the
QCD coupling constant and can be used as a non-perturbative tool to
investigate the strong coupling phenomena.

\newsection{Acknowledgements}

The author wishes to thank I. Kogan and A. Kovner for stimulating
discussions and P.P.A.R.C. for a research studentship.

\subsection*{Appendix}
\appendix
\newsection

The Gamma function and dimensional regularisation formulae and their
properties are widely published, e.g. \cite{10}, and so shall not be
given here in detail.  The dimensional regularisation formulae are
generally only quoted up to a numerator of $O(l^4)$, where $l$ is the
momentum integration variable.  In Appendices D and E the formulae for
numerators of $O(l^6)$ and $O(l^8)$ are required.  The necessary formulae are
given below.
\begin{eqnarray}
\int \frac{d^{2w}l}{(2\pi)^{2w}} \frac{l_\alpha l_\beta l_\mu l_\nu
l_\rho l_\sigma}{(l^2+M^2)^A} &=& \frac{1}{(4\pi)^w \Gamma(A) 8}
\left[ \delta_{\mu \nu} (\delta_{\rho \alpha} \delta_{\sigma \beta} +
\delta_{\sigma \alpha} \delta_{\rho \beta}) +  \delta_{\nu \sigma}
(\delta_{\mu \alpha} \delta_{\rho \beta} +
\delta_{\rho \alpha} \delta_{\mu \beta}) \right. \nonumber \\
& & + \delta_{\rho \sigma} (\delta_{\mu \alpha} \delta_{\nu \beta} +
\delta_{\nu \alpha} \delta_{\mu \beta}) + \delta_{\mu \rho} (\delta_{\nu \alpha} \delta_{\sigma \beta} +
\delta_{\sigma \alpha} \delta_{\nu \beta}) \nonumber \\
& & + \delta_{\nu \rho} (\delta_{\mu \alpha} \delta_{\sigma \beta} +
\delta_{\sigma \alpha} \delta_{\mu \beta}) + \delta_{\mu \sigma} (\delta_{\rho \alpha} \delta_{\nu \beta} +
\delta_{\nu \alpha} \delta_{\rho \beta}) \nonumber \\
& & + \left. \delta_{\alpha \beta} (\delta_{\mu \nu} \delta_{\rho \sigma} +
\delta_{\nu \rho} \delta_{\mu \sigma} + \delta_{\mu \rho} \delta_{\nu
\sigma}) \right] \frac{\Gamma(A-3-w)}{(M^2)^{A-3-w}}
\end{eqnarray}
\begin{eqnarray}
\int \frac{d^{2w}l}{(2\pi)^{2w}} \frac{l_\alpha l_\beta l_\gamma
l_\delta l^2 l^2}{(l^2+M^2)^A} &=&
\frac{1}{(4\pi)^w \Gamma(A)} \frac{63}{8} \left[\delta_{\alpha \beta}
\delta_{\gamma \delta} + \delta_{\beta \delta} \delta_{\alpha \gamma}
+ \delta_{\beta \gamma} \delta_{\mu \delta} \right]
\frac{\Gamma(A-4-w)}{(M^2)^{A-4-w} } \nonumber \\
\int \frac{d^{2w}l}{(2\pi)^{2w}} \frac{l_\alpha l_\beta l^2 l^2
l^2}{(l^2+M^2)^A} &=& \frac{1}{(4\pi)^w \Gamma(A)} \frac{315}{8}
\delta_{\alpha \beta} \frac{\Gamma(A-4-w)}{(M^2)^{A-4-w}}
\end{eqnarray}

\newsection

In this Appendix, $\Sigma$ of Section 4.1 is calculated.
\begin{equation}
\Sigma = -3g^2 C_2(G) q_m q_n \Delta^{-1}(q) \int \frac{d^4s}{(2\pi)^4}
\frac{(\delta_{mn} s^2 -s_m s_n)}{s^2(s^2+s_0^2)(s^2+2sq+q^2)}.
\label{b1}
\end{equation}
The loop integration over momentum requires the introduction of three
Feynman parameters, one of which is eliminated to give,
\begin{equation}
\int \frac{d^4s}{(2\pi)^4} \frac{(\delta_{mn}s^2 - s_m
s_n)}{s^2(s^2+s_0^2)(s^2 +2sq +q^2)} = \int_0^1 dx_1 \int_0^{1-x_1}
dx_2 \Gamma(3) \int \frac{d^4\tilde{s}}{(2\pi)^4} \frac{(\delta_{mn}
\tilde{s}^2 -\tilde{s}_m \tilde{s}_n)}{[\tilde{s}^2 +\tilde{s}_0^2x_2
+\tilde{M}^2]^3},
\label{r3}
\end{equation}
\begin{eqnarray}
\tilde{s}_0 &=& s_0 \nonumber \\
\tilde{s} &=& s + q(1-x_1-x_2) \nonumber \\
\tilde{M}^2 &=& q^2(1-x_1-x_2)(x_1+x_2)
\label{r4}
\end{eqnarray}
Using the standard dimensional regularisation formulae of Appendix A,
integration over the spatial and time-like components separtely gives,
\begin{eqnarray}
\Sigma &=& 3g^2 C_2(G) \frac{\Gamma(\epsilon)}{(4\pi)^2}\int_0^1 dx_1
\int_0^{1-x_1} dx_2 x_2^{-\frac{1}{2}} \nonumber \\
&=& \frac{4g^2 C_2(G)}{(4\pi)^2}\Gamma(\epsilon),
\label{r5}
\end{eqnarray}

\newsection

In this appendix, $\Pi$ of Section 4.2 is evaluated, where,
\begin{equation}
\Pi = -\frac{g^2}{2} C_2(G) \int \frac{d^4u}{(2\pi)^4} (2u_0+r_0)^2
\Delta^{-1}(r) ND.
\label{x5}
\end{equation}
The numerator and denominator, $N$ and $D$, are,
\begin{eqnarray}
N &=& \delta_{i'j'} \delta_{k'l'} (u^2\delta_{k'j'}-u_{k'}
u_{j'})((r+u)^2 \delta_{i'l'} - (r+u)_{i'} (r+u)_{l'}) \nonumber \\
D &=& \left[ u^2 (u^2+u_0^2) (r^2 +2ru +u^2) (r^2+ 2ru +u^2 +r_0^2
+2r_0u_0 +u_0^2)\right]^{-1}
\label{x6}
\end{eqnarray}
The method employed in the evaluation of $\Pi$ is similar to that used
for the evaluation of $\Sigma$ in section 3.1 although more complex.
It is necessary to introduce four Feynman parameters to use the
methods of dimensional regularisation.  Doing this and performing a
shift in momentum, the denominator can be written,
\begin{eqnarray}
\tilde{D} &=& \int_0^1 dx_1 \int_0^{1-x_1} dx_2 \int_0^{1-x_1-x_2}
dx_3 \Gamma(4) [\tilde{u}^2 + \tilde{u}_0^2(1-x_1-x_3) + \tilde{M}^2
]^{-4} \nonumber \\
\tilde{u} &=& u + r(1-x_1-x_2) \nonumber \\
\tilde{u}_0 &=& u_0 +\frac{r_0(1 -x_1-x_2 -x_3)}{(1 -x_1 -x_3)}
\nonumber \\
\tilde{M}^2 &=& r^2(1-x_1-x_2) +r_0^2(1-x_1-x_2-x_3) -r^2(1-x_1-x_2)^2
\nonumber \\
& & -\frac{r_0^2(1-x_1-x_2 -x_3)^2}{(1-x_1-x_3)}.
\label{r9}
\end{eqnarray}
The momentum in the numerator must also be shifted but before this is
written it is useful to consider what powers of $\tilde{u}_0$ and
$\tilde{u}_i$ in the numerator will contribute to the
$\beta$-function.  From the formulae of the Appendix we note that if
the numerator is quadratic in $\tilde{u}_0$ then the denominator must
be at most of power $\frac{3}{2}$.  i.e.
\begin{equation}
\int \frac{d \tilde{u}_0}{2\pi} \frac{\tilde{u}_0^2}{[\tilde{u}_0^2 +
\tilde{M}^2]^{\frac{3}{2}}} \sim \Gamma(\epsilon)
\label{r10}
\end{equation}
In order to have this second integration, because the initial
integrand has a denominator of power $4$ then the numerator must have
a component $\tilde{u}^2$.  i.e.
\begin{equation}
\int \frac{d^3 \tilde{u}}{(2\pi)^3} \frac{\tilde{u}^2
\tilde{u}_0^2}{[\tilde{u}^2+\tilde{u}_0^2 +\tilde{M}^2]^{\frac{3}{2}}}
\sim \frac{\tilde{u}_0^2}{[\tilde{u}_0^2 + \tilde{M}^2]^{\frac{3}{2}}}
\label{r11}
\end{equation}
Similarly, if there are no powers of $\tilde{u}_0$ in the numerator,
$\tilde{u}^4$ is required.  All odd powers of $\tilde{u}_0$ and
$\tilde{u}_i$ integrate to zero.  Therefore, only
terms quadratic and quartic in $\tilde{u}_i$ need be considered in the shifted
numerator, $\tilde{N}$.
\begin{equation}
\tilde{N} = 2\tilde{u}^2 \tilde{u}^2 + \tilde{u}^2 (r-2\tilde{r})^2
+ \tilde{u}_j \tilde{u}_l (-6\tilde{r}_j (r-\tilde{r})_l +
(r-\tilde{r})_j(r-\tilde{r})_l +\tilde{r}_j \tilde{r}_l) + ... 
\label{r12}
\end{equation}
The modification to the Coulomb is now written as,
\begin{eqnarray}
\Pi &=& -\frac{g^2}{2} C_2(G) \int \frac{d^4 \tilde{u}}{(2\pi)^4}
(4\tilde{u}_0^2 + (r_0 -\frac{2\tilde{r}_0}{(1-x_1 -x_3)})^2)
\Delta^{-1}(r) \nonumber \\
& & \int_0^1 dx_1 \int_0^{1-x_1} dx_2 \int_0^{1-x_1-x_2}
dx_3 \Gamma(4) \tilde{N} [\tilde{u}^2 + \tilde{u}_0^2(1-x_1-x_3)
+\tilde{M}^2]^{-4} \nonumber \\
&=& \Pi_{A1} +\Pi_{A2} +(r-2\tilde{r})^2(\Pi_{B1} + \Pi_{B2})
\nonumber \\
& & + (-6\tilde{r}_j (r-\tilde{r})_l + (r-\tilde{r})_j (r-\tilde{r})_l
+\tilde{r}_j \tilde{r}_l)(\Pi_{C1} + \Pi_{C2}),
\label{r13}
\end{eqnarray}
where $\Pi_{A1}$ to $\Pi_{C2}$ will be evaluated separtely.
\begin{eqnarray}
\Pi_{A1} &=& \int \frac{d^4 \tilde{u}}{(2\pi)^4} 8 \tilde{u}_0^2
\tilde{u}^2 \tilde{u}^2 [\tilde{u}^2 + \tilde{u}_0^2(1 -x_1 -x_3)
+\tilde{M}^2]^{-4} \nonumber \\
&=& \frac{15 \tilde{M}^2}{(4\pi)^2 \Gamma(4) (1-x_1
-x_3)^{\frac{3}{2}}} \Gamma(-1), \nonumber \\
\Pi_{A2} &=& 2 (r_0 -\frac{2\tilde{r}_0}{(1- x_1 -x_3)})^2 \int
\frac{d^4 \tilde{u}}{(2\pi)^4} \tilde{u}^2 \tilde{u}^2 [\tilde{u}^2 +
\tilde{u}_0^2 (1-x_1 -x_3) + \tilde{M}^2]^{-4} \nonumber \\
&=& \frac{15 (r_0 -\frac{2\tilde{r}_0}{(1- x_1 -x_3)})^2}{2(4\pi)^2
\Gamma(4) (1-x_1 -x_3)^{\frac{1}{2}}}\Gamma(\epsilon), \nonumber \\
\Pi_{B1} &=& \int \frac{d^4 \tilde{u}}{(2\pi)^4} 4 \tilde{u}_0^2
\tilde{u}^2 [\tilde{u}^2 + \tilde{u}_0^2(1 -x_1 -x_3)
+\tilde{M}^2]^{-4} \nonumber \\
&=& \frac{3}{(4\pi)^2 \Gamma(4) (1-x_1
-x_3)^{\frac{3}{2}}} \Gamma(\epsilon), \nonumber \\
\Pi_{B2} &=& 2 (r_0 -\frac{2\tilde{r}_0}{(1- x_1 -x_3)})^2 \int
\frac{d^4 \tilde{u}}{(2\pi)^4} \tilde{u}^2 [\tilde{u}^2 +
\tilde{u}_0^2 (1-x_1 -x_3) + \tilde{M}^2]^{-4} \nonumber \\
&=& {\rm finite} \nonumber \\
\Pi_{C1} &=& \int \frac{d^4 \tilde{u}}{(2\pi)^4} 4 \tilde{u}_0^2
\tilde{u}_j \tilde{u}_l [\tilde{u}^2 + \tilde{u}_0^2(1 -x_1 -x_3)
+\tilde{M}^2]^{-4} \nonumber \\
&=& \frac{\delta_{jl}}{(4\pi)^2 \Gamma(4) (1-x_1
-x_3)^{\frac{3}{2}}} \Gamma(\epsilon), \nonumber \\
\Pi_{C2} &=& 2 (r_0 -\frac{2\tilde{r}_0}{(1- x_1 -x_3)})^2 \int
\frac{d^4 \tilde{u}}{(2\pi)^4} \tilde{u}_j \tilde{u}_l [\tilde{u}^2 +
\tilde{u}_0^2 (1-x_1 -x_3) + \tilde{M}^2]^{-4} \nonumber \\
&=& {\rm finite}.
\label{r14}
\end{eqnarray}
Using the definition of the Gamma function in the Appendix,
$\Gamma(\epsilon) = -\Gamma(-1+\epsilon)$.  All finite terms are
ignored.  (\ref{r13}) is now written as,
\begin{eqnarray}
\Pi &=& -\frac{g^2}{2} \frac{C_2(G)}{(4\pi)^2}\Gamma(\epsilon)
\Delta^{-1} (r) \int_0^1 dx_1 \int_0^{1-x_1} dx_2 \int_0^{1-x_1 -x_2}
dx_3 \left[ -\frac{15\tilde{M}^2}{(1-x_1-x_3)^{\frac{3}{2}}} \right. \nonumber
\\
& & +
\frac{15}{2} \left(r_0-\frac{2\tilde{r}_0}{(1-x_1 -x_3)}\right)^2
(1-x_1 -x_3)^{-\frac{1}{2}} \nonumber \\
& & \left. + (1- x_1 -x_3)^{-\frac{3}{2}} \left(4(r -
\tilde{r})^2 - 12\tilde{r}(r-\tilde{r}) + 4\tilde{r} \tilde{r} \right) \right]
\label{r15}
\end{eqnarray}
Where spatial indices are omitted, summation is implied.  Considering
terms separately, again, execution of the Feynman parameter integrals
gives the results below.
\begin{eqnarray}
& & \int_0^1 dx_1 \int_0^{1-x_1} dx_2 \int_0^{1-x_1 -x_2}
dx_3 \frac{-15\tilde{M}^2}{(1- x_1 -x_3)^{-\frac{3}{2}}} =
-\frac{26}{7} r^2 - \frac{2}{3} r_0^2 \nonumber \\
& & \int_0^1 dx_1 \int_0^{1-x_1} dx_2 \int_0^{1-x_1 -x_2}
dx_3 \frac{15}{2} \left[r_0 -\frac{2\tilde{r}_0}{(1- x_1 -x_3)}
\right]^2 (1- x_1 -x_3)^{-\frac{1}{2}} = \frac{2}{3} r_0^2 \nonumber
\\
& & \int_0^1 dx_1 \int_0^{1-x_1} dx_2 \int_0^{1-x_1 -x_2}
dx_3 (1- x_1 -x_3)^{-\frac{3}{2}} \nonumber \\
& & \;\;\;\;\;\;\left[ 4(r-\tilde{r})^2 -12 \tilde{r}
(r- \tilde{r}) + 4 \tilde{r} \tilde{r} \right] = \frac{8}{21} r^2
\label{r16}
\end{eqnarray}
We finally write the modification to the Coulomb interaction by the
insertion of a transverse gluon loop as,
\begin{equation}
\Pi = -\frac{5}{3} \frac{g^2 C_2(G)}{(4\pi)^2} \Gamma(\epsilon)
\label{x4}
\end{equation}

\newsection

In this section one of the vertex corrections of Section 4.3 is
evaluated and found to be convergent.  Fig. 10 is represented by the
expression, 
\begin{equation}
-\frac{g^4}{2} f^{agf} f^{fbh} f^{ade} f^{hgc} \delta_{kl} (2r_0-q_0)
 \Delta^{-1}(q) \int \frac{d^4 s}{(2\pi)^4} (2p_0+s_0) ND,
\label{x18}
\end{equation}
where,
\begin{eqnarray}
N &=& (-s+q)_m\delta_{ij'} [(s-p)_{i'}\delta_{k'j}
+(p-q)_j\delta_{i'k'} + (2p+s-q)_{k'}\delta_{i'j}] \nonumber \\
& & ((p+s)^2\delta_{j'i'} -(p+s)_{j'} (p+s)_{i'})((s-q)^2 \delta_{k'm}
- (s-q)_{k'}(s-q)_{m}), \nonumber \\
D &=& [s^2 (p^2 +2ps +s^2) (s^2 +2ps +p^2 +s_0^2 +2p_0 s_0 + s_0^2)
(s^2 -2sq +q^2) \nonumber \\
& & (s^2 -2sq +q^2 + s_0^2 - 2s_0 q_0 +q_0^2)]^{-1}.
\label{x19}
\end{eqnarray}
Following the method of the previous sections, this denominator
requires five Feynman parameters to allow the use of the dimensional
regularisation formulae.  It can be written, after a shift in
momentum, as,
\begin{eqnarray}
\tilde{D} &=& \int_0^1 dx_1...\int_0^{1-x_1 -x_2 -x_3}dx_4 \Gamma(5)
[\tilde{s}^2 + \tilde{s}_0^2 (1-x_1-x_2-x_4) + \tilde{M}^2]^{-5},
\nonumber \\
\tilde{s} &=& s+\tilde{p}, \nonumber \\
\tilde{s}_0 &=& s_0 +\frac{\tilde{p}_0}{(1-x_1-x_2 -x_4)}, \nonumber
\\
\tilde{M}^2 &=& p^2(x_2+x_3) +q^2(1-x_1-x_2-x_3) +p_0^2x_3
+q_0^2(1-x_1 -x_2-x_3-x_4) \nonumber \\
& & -\tilde{p}^2
-\frac{\tilde{p}_0^2}{(1-x_1-x_2 -x_4)}.
\label{r20}
\end{eqnarray}
Inserting (\ref{r20}) into (\ref{x19}) and performing the same shift
in momentum throughout the integrand, in a similar manner to that used
in section 3.2 around (\ref{r10}) and (\ref{r11}) it is possible to
evaluate what powers of $\tilde{u}$ and $\tilde{u}_0$ in
the numerator contribute to the $\beta$-function.  First, we see that the term
linear in $\tilde{s}_0$ integrates to zero by symmetry.  Secondly, for
the final integration over time-like momentum to be divergent with a
constant numerator, a denominator with maximum power $\frac{1}{2}$ is
required.  For an initial denominator of power $5$ this implies a
numerator with terms of $O(\tilde{s}^6)$.  Writing out the terms in
$\tilde{N}$,
\begin{eqnarray}
\tilde{N} &=& - \tilde{s}_m [\tilde{s}_{i'} \delta_{k'j}+
\tilde{s}_{k'} \delta_{i'j}][\tilde{s}^2\delta_{ii'} - \tilde{s}_i
\tilde{s}_{i'}] [\tilde{s}^2 \delta{k'm} - \tilde{s}_{k'} \tilde{s}_m]
+O(\tilde{s}^5) \nonumber \\
&=& O(\tilde{s}^5)
\label{r21}
\end{eqnarray}
Therefore, Fig. 11 makes no contribution to the QCD $\beta$-function.

\newsection

In this section the second vertex correction of Section 4.3 is
calculated.  This corresponds to the evaluation of Fig. 11, which is
represented by the expression,
\begin{equation}
-\frac{g^4}{2} f^{ab'c'} f^{ade} f^{c'cg} f^{bb'g} \int \frac{d^4
s}{(2\pi)^4} (2s_0 +q_0)(2r_0-q_0) \delta_{kl} \Delta^{-1}(q) ND,
\label{x22}
\end{equation}
where,
\begin{eqnarray}
N &=& \delta_{j''i''} [(2p-s+q)_{i'}\delta_{jk''} + (-p-2q-s)_{k''}
\delta_{i'j} + (2s+q-p)_j \delta_{i'k''}] \nonumber \\
& & [(2s-p)_i \delta_{j'k'}
+(-s-p)_{k'} \delta_{ij'} + (2p-s)_{j'}\delta_{ik'}] \nonumber \\
& & [(-p+s)^2 \delta_{k''k'}
-(-p+s)_{k''}(-p+s)_{k'}] [s^2\delta_{j'j''} -s_{j'}s_{j''}] \nonumber
\\
& &[(q+s)^2
\delta_{i''i'} -(q+s)_{i''}(q+s)_{i'}], \nonumber \\
D &=& [s^2 (s^2+s_0^2) (p^2 -2ps +s^2) (p^2-2ps+s^2 +p_0^2 -2p_0s_0
+s_0^2) \nonumber \\
& & (q^2+2qs +s^2)(q^2+2qs+s^2 +q_0^2 +2q_0s_0 +s_0^2)]^{-1}.
\label{x23}
\end{eqnarray}
This denominator requires six Feynman parameters to allow the use of
the dimensional regularisation formulae.  It may be written as,
\begin{eqnarray}
\tilde{D} &=& \int_0^1 dx_1...\int_0^{1-x_1-x_2-x_3 -x_4} dx_5
\Gamma(6) [\tilde{s}^2 +\tilde{s}_0^2 (1-x_1 -x_3 -x_5) + \tilde{M}^2
]^{-6} \nonumber \\
\tilde{s} &=& s-\tilde{p} \nonumber \\
\tilde{s}_0 &=& s_0 -\frac{\tilde{p}_0}{(1-x_1-x_3-x_5)} \nonumber \\
\tilde{p} &=& p(x_1+x_2) -q(1-x_1-x_2-x_3-x_4) \nonumber \\
\tilde{p}_0 &=& p_0x_2 -q_0(1-x_1-x_2-x_3-x_4-x_5) \nonumber \\
\tilde{M}^2 &=& p^2(x_1+x_2) +p_0^2x_2 +q^2(1-x_1-x_2-x_3-x_4)
\nonumber \\
& & +q_0^2
(1-x_1-x_2-x_3-x_4-x_5) -\tilde{p}^2 -\frac{\tilde{p}_0^2}{(1-x_1-x_3-x_5)}
\label{r24}
\end{eqnarray}
Performing the same shift of momentum in the rest of the integrand,
(\ref{x22}) can be written as,
\begin{eqnarray}
& & -\frac{g^4}{2} f^{ab'c'} f^{ade} f^{c'cg} f^{bb'g}
\int_0^1...\int_0^{1-x_1- x_2-x_3-x_4} dx_5 \int \frac{d^4
\tilde{s}}{(2\pi)^4}(2r_0 -q_0) \delta_{kl} \nonumber \\
& & \left[ 2\tilde{s}_0 + \frac{2p_0x_2 - q_0(1-x_1 -2x_2 -x_3 -2x_4
-x_5)}{(1-x_1-x_3-x_5)}\right] \Delta^{-1}(q)
\tilde{N} \Gamma(6) \nonumber \\
& & [\tilde{s}^2 + \tilde{s}_0^2(1-x_1-x_3-x_5)
+\tilde{M}^2]^{-6}.
\label{r25}
\end{eqnarray}
The term linear in $\tilde{s}_0$ integrates to zero by symmetry.
Using methods described above, all the remaining terms in the
numerator are constant in $\tilde{s}_0$ and so only terms of
$O(\tilde{s}^8)$ will be infinite.  The momentum shifted numerator can
be written as,
\begin{equation}
\tilde{N} = 8\tilde{s}_i \tilde{s}_j \tilde{s}^6 +O(\tilde{s}^7).
\label{r26}
\end{equation}
Within (\ref{r25}), the integration over the loop momentum (using the
formulae of Appendix A) is,
\begin{equation} 8\int \frac{d^4\tilde{s}}{(2\pi)^4} \frac{\tilde{s}_i
\tilde{s}_j \tilde{s}^2 \tilde{s}^2 \tilde{s}^2}{[\tilde{s}^2 +
\tilde{s}_0^2(1-x_1-x_3-x_5) +\tilde{M}^2]^6} = 315
\frac{\delta_{ij}}{(4\pi)^2(1-x_1-x_3-x_5)^{\frac{1}{2}}}\Gamma(\epsilon).
\label{r27}
\end{equation}
The Feynman parameter integrals give,
\begin{equation}
\int_0^1 dx_1 \int_0^{1-x_1-x_2-x_3-x_4}dx_5 \left[\frac{2p_0x_2 - q_0
(1-x_1-2x_2 -x_3 -2x_4-x_5)}{(1-x_1-x_3-x_5)^{\frac{3}{2}}}\right] =
\frac{4}{945} (2p_0+q_0)
\label{r28}
\end{equation}
Therefore, substituting (\ref{r27}) and (\ref{r28}) in (\ref{r25}),
the vertex correction can be written as,
\begin{equation}
-\frac{2}{3} \frac{g^4}{(4\pi)^2} f^{ab'c'} f^{ade} f^{c'cg} f^{bb'g}
(2p_0+q_0) (2r_0 -q_0) \delta_{ij} \delta_{kl} \Delta^{-1}(q)
\Gamma(\epsilon).
\label{r29}
\end{equation}
Now, the four structure constants need to be shown to be equal to a
product of two.  As defined in (\ref{2}), $T^a$ and $f^{abc}$ are the
generators and structure constants of an SU(N) Lie group,
respectively.  In the adjoint representation,
\begin{equation}
f^{abc}=i(T^a)_{bc}
\label{r30}
\end{equation}
This allows relults known, e.g. \cite{9}, for the generators of an SU(N) Lie
group to be exploited.  For instance, the result ${\rm Tr} (T^a T^b
T^c) = \frac{N}{2}if^{abc}$ allows the evaluation,
\begin{equation}
f^{ab'c'} f^{ade} f^{c'cg} f^{bb'g} = -i(T^a)_{b'c'}(T^c)_{c'g}
(T^b)_{gb'} f^{ade} = -\frac{N}{2} f^{abc} f^{ade}.
\label{r31}
\end{equation}
Therefore, the Coulomb interaction between two transverse gluons with a
transverse gluon correction about one of the gluon - gluon - Coulomb
vertices is finally written as I$\Gamma$ where,
\begin{equation}
\Gamma = \frac{2}{3} \frac{g^2 C_2(G)}{(4\pi)^2}\Gamma(\epsilon).
\label{x32}
\end{equation}

\renewcommand{\footnotesize}{\small}

\noindent

\bigskip

{\renewcommand{\Large}{\normalsize}

\end{document}